# Smart Machine Learning for Solving the Full Electron-Nuclear Schrödinger Equation with a Quantum-Chemically Inspired Wavefunction Ansatz

Yaolong Zhang[1,*], Bin Jiang[2,*], and Hua Guo[1,*]

[1]*Department of Chemistry and Chemical Biology, Center for Computational Chemistry, University of New Mexico, Albuquerque, New Mexico 87131, USA*

[2]*State Key Laboratory of Precision and Intelligent Chemistry, Department of Chemical Physics, University of Science and Technology of China, Hefei, Anhui 230026, China*

*: corresponding authors, email: ylzhangch@unm.edu, bjiangch@ustc.edu.cn, hguo@unm.edu

**Abstract**

Directly solving the full electron-nuclear Schrödinger equation remains one of the grand challenges in quantum mechanics. Here, we propose a quantum-chemically motivated orbital-based neural network wavefunction ansatz with explicit many-body electron-electron correlation terms, with parameters depending on the entire electron-nuclear configuration, enabling an accurate description of strongly correlated and/or multi-reference electronic structures by a single Slater-determinant across a broad nuclear configuration space. To train this wavefunction, we further develop an efficient hybrid optimization strategy that combines the variational Monte Carlo technique with local-energy constraints, mitigating the intrinsic sampling bias of the former and considerably reducing the statistical error. These advances are embedded in a unified neural network framework, SchrödingerNet, enabling the determination of globally accurate potential energy surfaces with a single training for systems ranging from strongly correlated diatomic molecules, triatomic reactions containing conical intersections, to polyatomic multi-reference molecules. This work offers a practical machine-learning tool for solving the electron-nuclear Schrödinger equation and capturing full quantum effects beyond the Born-Oppenheimer approximation.

## Introduction

The Schrödinger equation (SE) provides a complete description of the microscopic world, yet its accurate solution for many-body systems remains computationally prohibitive due to the exponential growth of configurational complexity. In molecular systems, this challenge is almost always mitigated through the Born-Oppenheimer approximation (BOA)[1], which separates electronic and nuclear motion and reduces the problem to repeated solutions of the electronic SE at different nuclear geometries, followed by the solution of the nuclear SE on the resulting potential energy surface (PES). While such separation is intuitive based on mass disparity and highly efficient, there are many chemical processes that break the BOA, thus requiring difficult and expensive solutions of coupled multi-state problems[2-5].

Chemically accurate characterization of the electronic structure of molecular systems typically needs the inclusion of electron-electron correlation (EEC) in solving the electronic SE. This is often realized by using high-level post-Hartree-Fock electronic structure methods such as configuration interaction (CI)[6] and coupled cluster (CC)[7] approaches through explicit determinant expansions and exponential excitation operators, respectively. This typically comes with increasingly large multi-determinant expansions of the wavefunction, which grow combinatorially with the system size. In cases where a single reference is insufficient, multi-reference methods must be used.

Furthermore, the solution of the nuclear SE requires accurate global PESs, which can only be constructed by repeating the expensive electronic structure calculations across a high-dimensional nuclear configuration space, further compounding the computational intensity[8, 9]. This problem is particularly acute if multiple electronic states are involved[5].

Recent advances in machine learning (ML) have opened new opportunities for electronic structure theory[10-12]. In particular, neural-network (NN) based wavefunction ansatzes[10, 13-40] combined with variational Monte Carlo (VMC)[41] techniques have demonstrated impressive accuracy for solving the electronic SE at fixed nuclear geometries for various systems, often with more favorable scaling than traditional high-level basis-set methods[42]. Since the NN wavefunction is typically not built from single-electron orbitals, its multi-reference nature can be captured by directly encoding many-body correlation through NNs. However, these pioneering approaches have not fully taken advantage of known properties of molecular systems, leaving ample room for improvement. We note in passing that constructing a global PES with an NN-based wavefunction method still requires repeated VMC-based optimizations at different nuclear configurations[43], although parameter-sharing schemes have been proposed to partially reuse information across geometries that can improve initialization and accelerate convergence of the optimization procedure[24-26, 44].

A conceptually different strategy is to directly solve the full SE rather than treating electronic and nuclear problems separately[45-47]. Such a strategy is particularly attractive for problems beyond the BOA. Our recently proposed SchrödingerNet framework,[45] which represents an important step in this direction, introduces an NN based total wavefunction ansatz that depends explicitly on both electronic and nuclear coordinates. In particular, the training is carried out in the Cartesian space, in which kinetic energy operators (Laplacians) can be readily evaluated and potential energy operators are local. The system is variationally brought to convergence through ML to the combined ground state. Importantly, the entire PES can be readily obtained by removing the nuclear kinetic energy operator from the total Hamiltonian. In our proof-of-concept work[45], SchrödingerNet was demonstrated to yield accurate potential energy curves for simplest diatomic molecules through a single wavefunction optimization. However, extending this framework to larger and strongly correlated and/or multi-reference systems demands a more expressive wavefunction representation and a more efficient optimization scheme.

Although the expressivity of an NN can generally be increased by brute force, such as the enlargement of the parameter space, it is more advantageous to design smart ansatzes based on intrinsic physics of the system of interest. In this work, we introduce a conceptually novel wavefunction ansatz inspired by quantum-chemistry. In this compact form, the electronic wavefunction is expressed as a single Slater-determinant built from linear combinations of many-electron atomic orbitals, but with their orbital coefficients and exponents dynamically adaptive to the electronic and nuclear coordinates of the entire system. In addition, an additional term is added to explicitly capture EEC effects. The key designing principle here is to retain the existing physical concepts in ab initio quantum chemistry, but to take advantage of the flexibility and efficiency of ML approaches. Further, we propose a hybrid training strategy that combines VMC energy minimization with local-energy consistency constraints[45], enabling fast and even convergence of the total wavefunction over a broad nuclear configuration space by enhancing sampling of nuclear configurations far from equilibrium. This improved SchrödingerNet setup allows a uniform characterization of equilibrium structures, global reactive pathways, nonadiabatic regions, and multi-reference and/or strongly correlated polyatomic molecules within a single wavefunction training using a compact single-determinant representation, as demonstrated by several challenging examples.

## Methods

SchrödingerNet[45] targets a direct solution of the full non-relativistic SE for a molecular system containing $N_n$ nuclei and $N_e$ electrons. In atomic units, the Hamiltonian is given in atomic units by,

$$\hat{H} = -\sum_{a=1}^{N_n} \frac{1}{2m_a} \nabla_{\boldsymbol{R}_a}^2 - \frac{1}{2}\sum_{i=1}^{N_e} \nabla_{\boldsymbol{r}_i}^2 + \sum_{i>j}^{N_e} \frac{1}{|\boldsymbol{r}_i - \boldsymbol{r}_j|} - \sum_{i=1}^{N_e}\sum_{a=1}^{N_n} \frac{Z_I}{|\boldsymbol{r}_i - \boldsymbol{R}_a|} + \sum_{a>b}^{N_n} \frac{Z_a Z_b}{|\boldsymbol{R}_a - \boldsymbol{R}_b|} \quad , \tag{1}$$

where $\boldsymbol{r}_i$ and $\boldsymbol{R}_a$ denote the Cartesian coordinates of the $i$th electron and $a$th nucleus, respectively, and $m_a$ and $Z_a$ are the corresponding nuclear masses and charges. The stationary states of the system satisfy the time-independent SE,

$$\hat{H}\psi(\boldsymbol{r}, \boldsymbol{R}) = E\psi(\boldsymbol{r}, \boldsymbol{R}), \tag{2}$$

where the total wavefunction $\psi(\boldsymbol{r}, \boldsymbol{R})$ depends explicitly on both electronic and nuclear configurations. In the BOA, the total wavefunction is factorized into electronic and nuclear components, where the former obtained from solving the electronic SE depends parametrically on nuclear coordinates, while the latter depending solely on nuclear positions is obtained by solving the nuclear SE subsequently. Although this factorization has proven highly successful in many contexts, it breaks down when non-adiabatic effects become dominant[2-5, 47].

For our purpose, we adopt a non-BO wavefunction ansatz in SchrödingerNet, in which electronic and nuclear degrees of freedom are treated on an equal footing[45]. Specifically, we represent the total wavefunction as

$$\psi(\boldsymbol{r},\boldsymbol{R}) = \psi^{(e)}(\boldsymbol{r},\boldsymbol{R})\psi^{(n)}(\boldsymbol{r},\boldsymbol{R}), \tag{3}$$

where the electronic and nuclear components, $\psi^{(e)}(\boldsymbol{r},\boldsymbol{R})$ and $\psi^{(n)}(\boldsymbol{r},\boldsymbol{R})$, both depend explicitly on electronic and nuclear coordinates, thus allowing full coupling of all degrees of freedom. A schematic of the overall design of joint electron-nuclear wavefunction ansatz is displayed in Fig. 1.

**Electronic Wavefunction Ansatz**

To satisfy the Pauli principle for the fermionic electrons, $\psi^{(e)}(\boldsymbol{r},\boldsymbol{R})$ is conveniently expressed as a Slater determinant,

$$\psi^{(e)}(\boldsymbol{r},\boldsymbol{R}) = det[\varphi_o(\boldsymbol{r}_i;\boldsymbol{r},\boldsymbol{R})], \tag{4}$$

which enforces antisymmetry under permutation of electronic positions. Here, $\varphi_o(\boldsymbol{r}_i;\boldsymbol{r},\boldsymbol{R})$ denotes the $o$th molecular orbital (MO) function evaluated at the electronic position $\boldsymbol{r}_i$. Each electron is associated with a MO, much like the unrestricted Hartree-Fock scheme. Although the MO concept is borrowed from conventional quantum chemistry, it should be stressed that the MOs in eq. (4) depend explicitly, rather than parametrically, on all electronic ($\boldsymbol{r}$) and nuclear coordinates ($\boldsymbol{R}$).

In traditional quantum chemistry[48], MOs are typically constructed as linear combinations of atom orbitals (LCAO):

$$\tilde{\varphi}(\boldsymbol{r}_i\colon \boldsymbol{R}) = \sum_a^{N_n} \sum_{n=1}^{N} \sum_{l=0}^{L} \sum_{m=-l}^{l} c_{nlm} \chi_{nlm}(r_{ia})\, Y_{lm}(\hat{\boldsymbol{r}}_{ia}), \tag{5}$$

where $\chi_{nlm}(r_{ia})$ is a radial function (e.g., Slater- or Gaussian-type) and $Y_{lm}(\hat{\boldsymbol{r}}_{ia})$ denote spherical harmonics or their Cartesian equivalents with the quantum number $l$ specifying the atomic orbital type ($s$, $p$, $d$, etc.). The expansion coefficients ($c_{nlm}$) are parameters optimized variationally through the self-consistent field (SCF) procedure[48]. $N$ and $L$ denote the principal and angular-momentum quantum numbers, respectively. While this atom-center representation efficiently captures the dominant electron-nuclear interactions, it does not optimally describe EEC effects between electrons, which are vital to achieve chemical accuracy. Consequently, EEC is often treated approximately through extensive CI expansions, which is very inefficient. We further emphasize that the LCAO-MOs depend parametrically on the nuclear coordinates, denoted by : in eq. (5).

Obviously, this problem is avoided in non-basis-set methods such as quantum Monte Carlo approaches, where the wavefunction does not need be expressed as LCAO and parameters are optimized directly by a stochastic algorithm, incorporating EEC effects through approaches such as backflow transformations[41]. In the same spirit, our ML approach also optimizes the parameters of an NN wavefunction ansatz directly. This flexibility enables the use of configuration-dependent coefficients in our ansatz, which can be written in terms of atom-centric orbital functions:[45]

$$\tilde{\phi}_{nlm}(\boldsymbol{r}_i, \boldsymbol{R}_a; \boldsymbol{r}, \boldsymbol{R}) = IA_n^c(\boldsymbol{r}, \boldsymbol{R})\, exp(-(IA_n^\alpha(\boldsymbol{r}, \boldsymbol{R}))^2 * r_{ia})\, Y_{lm}(\hat{\boldsymbol{r}}_{ia}). \tag{6}$$

Importantly, the orbital coefficients $IA_n^c(\boldsymbol{r}, \boldsymbol{R})$ and exponents $IA_n^\alpha(\boldsymbol{r}, \boldsymbol{R})$ depend on the electronic and nuclear positions and are generated by message-passing neural networks (MPNNs)[49, 50], thereby making each orbital a many-electron function rather than a conventional one-electron function. Here, the notations "*I*" and "*A*" refer to the central node (electron *i*) and a neighboring node (atom *a*), respectively, while "*IA*" denotes the edge connecting them. The superscripts *c* and *α* distinguish orbital coefficients and exponents, while the subscript *n* represents different radial functions (or channels, in ML terminology). Similar notations are adopted throughout this work.

A key innovation in this work is the introduction of an explicit EEC term in the electronic wavefunction ansatz:

$$I_{nlm}^{ee} = I_{nlm} \sum_{j \neq i}^{N_e} IJ_n^c(\boldsymbol{r}, \boldsymbol{R}) \, exp\left(-IJ_n^\alpha(\boldsymbol{r}, \boldsymbol{R})/(1.0 + r_{ij})\right), \quad (7)$$

where the radial decay function adopts a form designed to satisfy the electron-electron cusp condition at $r_{ij} = 0$. Similarly, here $IJ_n^c(\boldsymbol{r}, \boldsymbol{R})$ and $IJ_n^\alpha(\boldsymbol{r}, \boldsymbol{R})$ are also configuration-dependent quantities and generated by MPNNs, while $I_{nlm}$ denotes node equivariant features generated by MPNNs. This explicit EEC term is then coupled with the many-electron orbital functions in eq. (6) through a tensor-product construction to mimic the shielding effect of the other electrons:

$$\phi_{nlm}(\boldsymbol{r}_i, \boldsymbol{R}_a; \boldsymbol{r}, \boldsymbol{R}) = \sum_{l_1 m_1, l_2 m_2} C_{l_1 m_1, l_2 m_2, lm} \tilde{\phi}_{n l_1 m_1}(\boldsymbol{r}_i, \boldsymbol{R}_a; \boldsymbol{r}, \boldsymbol{R}) I_{n l_2 m_2}^{ee}. \quad (8)$$

Here, $C_{l_1 m_1, l_2 m_2, lm}$ are the corresponding Clebsch-Gordan coefficients. We emphasize that the orbital functions in eq. (8) are no longer purely atom-centric, because of the $r_{12}$ terms introduced in eq. (7). These EEC-encoded basis functions thus have the potential to substantially improve the capturing of EEC in SchrödingerNet, as demonstrated below. We emphasize that it is very difficult to directly introduce such physically important terms in conventional quantum chemistry theory because of the Coulomb singularity at $r_{12} = 0$. Yet in our approach, it is straightforward and does not have a large effect on numerical efficiency.

Finally, MOs in the Slater determinant are constructed in SchrödingerNet through linear combinations of the orbital functions defined in Eq. (8), with the corresponding mixing coefficients $\boldsymbol{c}_{o,nlm}(\boldsymbol{r},\boldsymbol{R})$ represented by MPNNs[51],

$$\varphi_o(\boldsymbol{r}_i;\boldsymbol{r},\boldsymbol{R}) = \sum_n \sum_{l,m} \sum_{a=1}^{N_n} \boldsymbol{c}_{o,nlm}(\boldsymbol{r},\boldsymbol{R}) \phi_{nlm}(\boldsymbol{r}_i,\boldsymbol{R}_a;\boldsymbol{r},\boldsymbol{R}). \quad (9)$$

Notably, the MOs are single-electron orbitals with awareness of all electronic and nuclear coordinates through their configuration-adaptive expansion coefficients.

We further emphasize that although the electronic wavefunction formally consists of a single Slater determinant, it is not tied to a single-reference Hartree-Fock representation in the conventional sense. This is because multi-reference character is encoded functionally through configuration-adaptive orbital coefficients and the explicit EEC term, rather than through an expansion over multiple determinants. Furthermore, mixing orbitals with different angular momenta (*s*, *p*, *d*, etc.) up to a maximum angular momentum $L$ naturally incorporates orbital hybridization into the ansatz, enabling an accurate description of various chemical interactions. As a result, a single determinant should be sufficient to describe electronic states whose characters may vary across the nuclear configuration space, even in situations where different regions would conventionally require distinct CI expansions. To the best of our knowledge, such an intrinsic multi-reference character within a single-determinant framework has not been validated in any previous wavefunction ansatz.

**Nuclear Wavefunction Ansatz**

The nuclear wavefunction $\psi^{(n)}(\boldsymbol{r},\boldsymbol{R})$ is constructed to provide a flexible, permutation-invariant description of nuclear correlations while remaining fully coupled to the electronic degrees of freedom. The design also ensures the correct boundary behavior: the wavefunction vanishes when any two nuclei approach each other or when all nuclei are infinitely separated (In this work we focus on bound states.). To satisfy these conditions, we adopt the following functional form,

$$\psi^{(n)}(\boldsymbol{r},\boldsymbol{R}) = \left(\sum_n exp\left(\sum_{a\neq b}^{N_n} -\frac{\left(AB_n^{\alpha 1}(\boldsymbol{r},\boldsymbol{R})\right)^2}{R_{ab}}\right)\sum_{a\neq b}^{N_n} AB_n^{c}(\boldsymbol{r},\boldsymbol{R})\, exp\left(-\left(AB_n^{\alpha 2}(\boldsymbol{r},\boldsymbol{R})\right)^2 R_{ab}\right)\right)^2, \tag{10}$$

where $R_{ab}$ denotes the inter-nuclear distance between nuclei $a$ and $b$. The nuclear wavefunction is represented as a linear combination of the basis functions where $AB_n^{c}(\boldsymbol{r},\boldsymbol{R})$, $AB_n^{\alpha 1}(\boldsymbol{r},\boldsymbol{R})$ and $AB_n^{\alpha 2}(\boldsymbol{r},\boldsymbol{R})$ are generated by a MPNN model. It is worth noting that although $\psi^{(n)}(\boldsymbol{r},\boldsymbol{R})$ is formally expressed as a function of nuclear coordinates, it implicitly depends on the electronic coordinates through the message-passing layers used to generate $AB_n^{c}(\boldsymbol{r},\boldsymbol{R})$, $AB_n^{\alpha 1}(\boldsymbol{r},\boldsymbol{R})$ and $AB_n^{\alpha 2}(\boldsymbol{r},\boldsymbol{R})$.

Specifically, our recently developed node equivariant message-passing (NEMP) architecture[51] is the chosen MPNN used throughout the total wavefunction, which has shown to be much more efficient and memory-efficient than conventional edge-based equivariant message-passing architectures when learning PESs. The edge outputs (e.g., *IJ*, *AB* and *IA*) and node outputs (e.g., *I*, *J*, and *A*) of NEMP are subsequently used to generate configuration-dependent orbital coefficients, orbital exponents, and other parameters required by the wavefunction ansatz.

**Loss function**

With a well-defined wavefunction ansatz, the design of an appropriate loss function is crucial for achieving stable and efficient optimization. A natural choice is based on VMC and using the expectation value of the Hamiltonian as the loss function,

$$\mathcal{L} = \frac{\langle\psi|\hat{H}|\psi\rangle}{\langle\psi|\psi\rangle}, \tag{11}$$

which has been widely used to yield highly accurate target energies for the electronic SE[17, 41]. Although the VMC objective provides an unbiased variational estimate of the total energy, its sampling distribution is proportional to $|\psi|^2$. Consequently, the contributions from nuclear geometries far from equilibrium contributes only marginally to the variational objective, resulting in unstable and inaccurate energy predictions in those regions.

To address this issue, we introduce an additional local-energy based loss function term[45],

$$\mathcal{L} = \frac{\langle\psi|\hat{H}|\psi\rangle}{\langle\psi|\psi\rangle} + \omega \sum_k (E_L(\{\boldsymbol{r}, \boldsymbol{R}\}_k) - E_{VMC})^2. \tag{12}$$

In this hybrid loss function, the first term estimates the total energy variationally via VMC, while the second local-energy term minimizes the difference between the local energy ($E_L$) and the reference VMC energy ($E_{VMC}$) over all sampled configurations—a prerequisite imposed by the SE—thereby mitigating the sampling bias induced by the $|\psi|^2$ distribution. $\{\boldsymbol{r}, \boldsymbol{R}\}_k$ denotes *k*th electron-nuclear configuration. The weighting factor $\omega$ balances the global energy optimization against the local energy consistency constraint.

In practice, the hybrid optimization benefits from two complementary sampling objectives. First, the regular VMC sampler can provide an estimate of the reference energy $E_{VMC}$, which serves as the target of the local-energy loss function. Second, a more refined sampling can be done around selected nuclear geometries that are underrepresented by the VMC distribution. Because this second sampler is used only to identify poorly described configurations, it needs not follow the VMC distribution. Together, the two procedures combine global variational energy estimation with targeted refinement of underrepresented regions, thereby promoting a more uniform accuracy across nuclear configuration space.

Numerically, the dominant computational cost arises from evaluating the Slater determinant and the kinetic-energy Laplacian. For a fixed number of determinants, evaluation of the Slater matrix scales as $O(N_e^3)$, whereas a naive coordinate-wise calculation of the Laplacian would lift the upper-bound scaling to $O((N_e+N_n)N_e^3)$. To speed up the optimization procedure, we employ the forward-Laplacian algorithm[33] provided by the folx package[52], which reduces the computational overhead relative to a naive Hessian-based evaluation. The practical scaling is therefore expected to lie somewhere between $O(N_e^3)$ and $O(N_e^4)$, which is much better than existing high-level ab initio methods.

## Results

### Effect of the explicit EEC term

We first validate the effect of the explicit EEC term in our wavefunction ansatz. We take F and Cl atoms as examples with multiple electrons and strong EECs, but free of nuclear coordinates. The learning curves of electronic energies of F and Cl obtained with and without the explicit EEC term included in the wavefunction are compared in Figs. 2 (a-d). For both atoms, the physically inspired EEC-explicit ansatz converges the wavefunction optimization more rapidly and exhibits considerably smaller relative errors in the converged energies. In principle, even without an explicit EEC term, the wavefunction itself can encode EEC implicitly through the configuration-dependent orbital coefficients and exponents. In practice, however, this may require substantially more epochs to achieve a comparable accuracy. As shown in Table I, the EEC-explicit ansatz reaches lower electronic energies than its EEC-free counterpart, much closer to the FermiNet benchmarks[19, 53]. Note that the root-mean-square deviation of the sampled local energies from the corresponding VMC energy is approximately $10^{-3}$ of the absolute total energy, which is approximately one order of magnitude smaller than the results of FermiNet[28, 54]. These results suggest that the explicit inclusion of the EEC term in our wavefunction ansatz significantly improves the description of the electronic structure of these systems. Notably, the improvement is particularly pronounced for Cl, indicating that the EEC term plays a more critical role as the complexity of electronic structure increases.

In addition, to assess the capability of a single Slater-determinant in describing the EEC effects, we compare in Fig. 2 (e-f) the learning curves of the F atom case with one, four, and eight-determinants, while keeping all other components of the wavefunction unchanged. Increasing the number of determinants in the ansatz yields neither a faster convergence behavior nor a better final variational energy than the single-determinant ansatz. These results strongly suggest that, at the present level of accuracy, the configuration-adaptive orbitals already provide sufficient flexibility within a single determinant.

**Multi-reference and strongly correlated diatomic molecules**

We next show that performance of SchrödingerNet in mapping accurate global potential energy curves (PECs) for $Li_2$ and $Be_2$ through a single electron-nuclear wavefunction optimization. $Li_2$ is a weakly bound covalent molecule, and its dominant electronic character changes continuously as the bond is stretched from the equilibrium geometry toward dissociation[55]. It therefore provides a useful benchmark for determining whether a single-determinant, configuration-adaptive orbital-based representation can describe both the bonding and dissociation regimes. As shown in Fig. 3(a), a SchrödingerNet model trained with a maximum orbital angular momentum $L$=1 successfully generates a PEC in excellent agreement with the Free Complement Theory (FCT) reference[55] in the entire internuclear distance range, accurately reproducing the well depth, equilibrium bond length, and dissociation behavior. Interestingly, restricting the orbital components to $L = 0$ (or equivalently using solely *s* orbitals) produces a substantially shallower well and shifts the minimum toward a longer bond length. Nevertheless, the $L = 0$ curve remains close to the reference in the asymptotic region, suggesting that its principal deficiency lies in the description of molecular bonding rather than the separated-atom limit. This behavior reflects the fact that low-lying 2*p* orbitals deeply participate in $Li_2$ bonding[56, 57], despite its ostensibly single bond character. This result underscores the importance of incorporating *s*–*p* mixing and directional flexibility into the proposed wavefunction ansatz for an accurate and balanced description of the bonding and dissociation regimes.

$Be_2$ provides a substantially more challenging benchmark. Despite containing only eight electrons, its weak bond results from a delicate balance of correlation effects and multi-reference nature arising from the near-degeneracy of the 2*s* and 2*p* orbitals[58-60]. As the atoms approach each other, pronounced configurational mixing and orbital rehybridization produce a characteristic change in slope near an internuclear distance of 6.3 bohr, appearing as a shoulder. As shown in Fig. 3(b), SchrödingerNet reproduces not only the $Be_2$ equilibrium position and well depth but also this subtle shoulder structure determined by experimental measurements[60]. Because the interaction energy is small, the potential is displayed over a narrow energy window that magnifies residual deviations. The close agreement with the experimentally derived potential[58] under this stringent scale therefore provides an important stress test of the accuracy of the proposed wavefunction ansatz.

This system also demonstrates the effectiveness of the new hybrid training strategy. As shown in Fig. 3(b), even with the EEC-explicit wavefunction ansatz, VMC barely achieves a reasonable accuracy near equilibrium, where nuclear configurations are sampled with relatively high probabilities, but its errors increase drastically in weakly sampled regions, yielding a highly oscillating dissociation behavior. In contrast, the hybrid loss function results in a PEC in excellent agreement with the experimentally derived curve over the full coordinate range. These results suggest that the wavefunction design and hybrid optimization algorithm play complementary roles: the former provides the flexibility required to represent varying electronic structures, whereas the latter ensures that this flexibility is optimized uniformly across nuclear configuration space.

**Multidimensional potential energy surfaces of polyatomic systems**

**$H+H_2$ reaction**

The $H+H_2$ reaction is the simplest chemical reaction and has long served as a benchmark for revealing fundamental phenomena in molecular reaction dynamics[61]. Particularly, at equilateral triangular geometries, the ground and first excited electronic states become degenerate, giving rise to a conical intersection and associated geometric-phase effects[62, 63]. This system thus offers a stringent test of whether our approach can generate an accurate three-dimensional PES covering not only the reaction channels, but also the conical-intersection region.

Fig. 4(a) shows the energy landscape obtained by SchrödingerNet with the $H_1$-$H_2$-$H_3$ angle fixed at 180°, as a function of the $H_1$-$H_2$ and $H_2$-$H_3$ distances. The predicted PES agrees closely with a very recent PES based on multi-reference configuration interaction (MRCI) calculations[64] and precisely reproduces the reaction pathway, the barrier height, and the overall topology of the reactive landscape. Within the energy range of -1.672 to -1.590 a.u. shown in Fig. 4(a), the root-mean-square deviation between the SchrödingerNet predictions and the MRCI energies is around $1.3\times10^{-4}$ a.u. or 3.5 meV. To examine the conical-intersection region more directly, Fig. 4(b) compares the predicted energy profile along a path passing through this region with high-level MRCI results[64]. The $D_{3h}$ conical intersection is clearly seen in the figure as a cusp of the potential. MRCI energies exhibit an approximately constant positive offset above the variational SchrödingerNet energies, presumably owing to the use of a finite basis-set in the former. Interestingly, after applying a constant shift to MRCI energies, the Shifted-MRCI curve coincides with the SchrödingerNet curve along the entire path, highlighting the accuracy of SchrödingerNet for describing the conical interaction region of the PES.

**Umbrella inversion of $NH_3$**

We next investigate the umbrella inversion of $NH_3$, which is a textbook example quantum mechanical tunneling. SchrödingerNet correctly describes the ground vibrational state of $NH_3$ with its probability density distribution symmetrically populates the two $C_{3v}$ pyramidal minima of the double-well potential of the ground electronic state, as is clearly shown in Fig. 5. Specifically, the classical equilibrium geometry on the PES, characterized by the symmetric N-H bond distance ($R_{NH}$) and the umbrella angle ($\chi$) as illustrated in Fig. 5a, lies at $R_{NH}$ = 1.91 bohr and $\chi$ = 111.5° (or symmetrically 68.5°). A planar saddle point for inversion exists at $R_{NH}$ = 1.88 bohr and $\chi$ = 90°. These geometries are nearly identical to those obtained by the high-level explicitly correlated coupled cluster singles, doubles, and perturbative triples (CCSD(T)-F12a) calculations.[65] The calculated inversion barrier of 0.00829 a.u. is also very close to that obtained with CCSD(T)-F12a (0.00814 a.u.).[65] These results highlight that nuclear wavefunction can be extracted directly and accurately from the SchrödingerNet framework, without the need to solve the nuclear SE based on a trained PES, as required in conventional BOA treatments.

**Cyclobutadiene: a test for a larger multireference molecule**

Finally, we apply SchrödingerNet to cyclobutadiene ($C_4H_4$), a larger polyatomic molecule with strong multi-reference character[66]. Cyclobutadiene has long served as a demanding benchmark for electronic-structure methods, because its rectangular equilibrium structure and square saddle point possess qualitatively different electronic characters[18]. Here, we calculate the electronic energies at the previously reported geometries of the minimum and the saddle point[45], and the corresponding learning curves are shown in Fig. 6. Clearly, both calculations converge rapidly and reach rather stable energies after approximately 25,000 epochs, indicating the robustness of SchrödingerNet despite the increased electronic complexity of the system. In comparison, the FermiNet calculations for this system converge much more slowly, with their learning curves only beginning to level off at 200,000 epochs[53]. These results highlight the effectiveness of integrating quantum-chemical concepts with neural-network flexibility to achieve accurate wavefunction representations and rapid convergence.

The predicted automerization barrier height of cyclobutadiene is also compared with earlier results obtained by representative multi-reference, coupled-cluster, and NN wavefunction methods,[18] including complete active space SCF (CASSCF), coupled cluster singles, doubles, triples and quadruples (CCSDTQ), PauliNet, and FermiNet, as summarized in Table II. Our approach yields a barrier height of 9.60 kcal/mol, which is closest to the most reliable CCSDTQ value. This indicates that the proposed wavefunction ansatz in SchrödingerNet accommodates the substantial electronic reorganization accompanying the structural transformation, without requiring separate determinant expansions tailored to the two stationary points.

**Discussion and Conclusion**

The current prevailing protocol for solving many-body problems in molecular systems starts with the BOA, which stipulates the initial solution of electronic SE at fixed nuclear geometries, followed by solution of the nuclear SE on one or more PESs. The parametric dependence of the electronic wavefunction on nuclear coordinates is on the one hand a blessing as it substantially reduces the dimensionality of the problem, but on the other hand can be a curse because of prevalent nonadiabatic effects.

Within the BOA, electron correlation and multi-reference character remain two central challenges in modern electronic-structure theory. In the Hartree-Fock framework, molecular orbitals constructed from atom-centered one-electron functions do not explicitly depend on the positions of other electrons, while a single reference may be inadequate when several electronic configurations contribute substantially. These problems are particularly acute for highly excited states of energetic molecules, where multiple near-degenerate electronic states are coupled. How to efficiently and accurately account for correlation and multi-reference character might require drastically different approaches. Recent progress in ML solution of the electronic SE has shown great promises of these non-traditional schemes, but few have ventured beyond the BOA.

In the meantime, the concept of MO based Slater determinants widely used in conventional electronic structure theory remain valuable in ML solution of SE. A smart strategy is to encode these useful and time-tested concepts in efficient ML solution of SE, preferably without the BOA. To this end, we design a quantum-chemically inspired, configuration-adaptive orbital ansatz incorporating explicit EEC terms, coupled with a hybrid optimization strategy based on VMC energy minimization and local-energy consistency constraints. This approach retains useful conventional quantum-chemical concepts while exploiting the flexibility and numerical efficiency of ML to directly solve the full electron-nuclear SE without invoking the BOA. The configuration-dependent orbital parameters and explicit EEC terms accommodate correlated and multi-reference electronic structures within a compact single-determinant representation, while the hybrid optimization improves accuracy in nuclear configurations that are weakly sampled by conventional VMC. Applications to strongly correlated atoms, weakly bound diatomic molecules, reactive systems, and multi-reference larger molecules demonstrate close agreement with high-level benchmark results. These developments establish SchrödingerNet as both a foundational and practical tool for extending full electron–nuclear quantum mechanical calculations to increasingly complex molecular systems.

Although the present work focuses on the ground state, orthogonality-constrained optimization and other techniques could extend the framework to excited electron-nuclear states[28, 36], offering a promising route toward a fully quantum description of nonadiabatic processes. Furthermore, the orbital-based ansatz offers opportunities for localized-orbital approximations and sparse Slater-matrix representations, which could reduce computational costs and facilitate applications to larger molecular systems. In sum, physically inspired frameworks such as SchrodingerNet offers smart ML treatments of quantum many-body problems that is central to chemistry.

**Acknowledgements:** This work at UNM is supported by the NSF (Grant No. CHE-2306975). The work at USTC is supported by National Natural Science Foundation of China (22325304) and the robotic AI-Scientist platform of Chinese Academy of Sciences. We acknowledge the use of OpenAI Codex to assist with the preparation of the schematic illustrations in this work.

## FIGURES

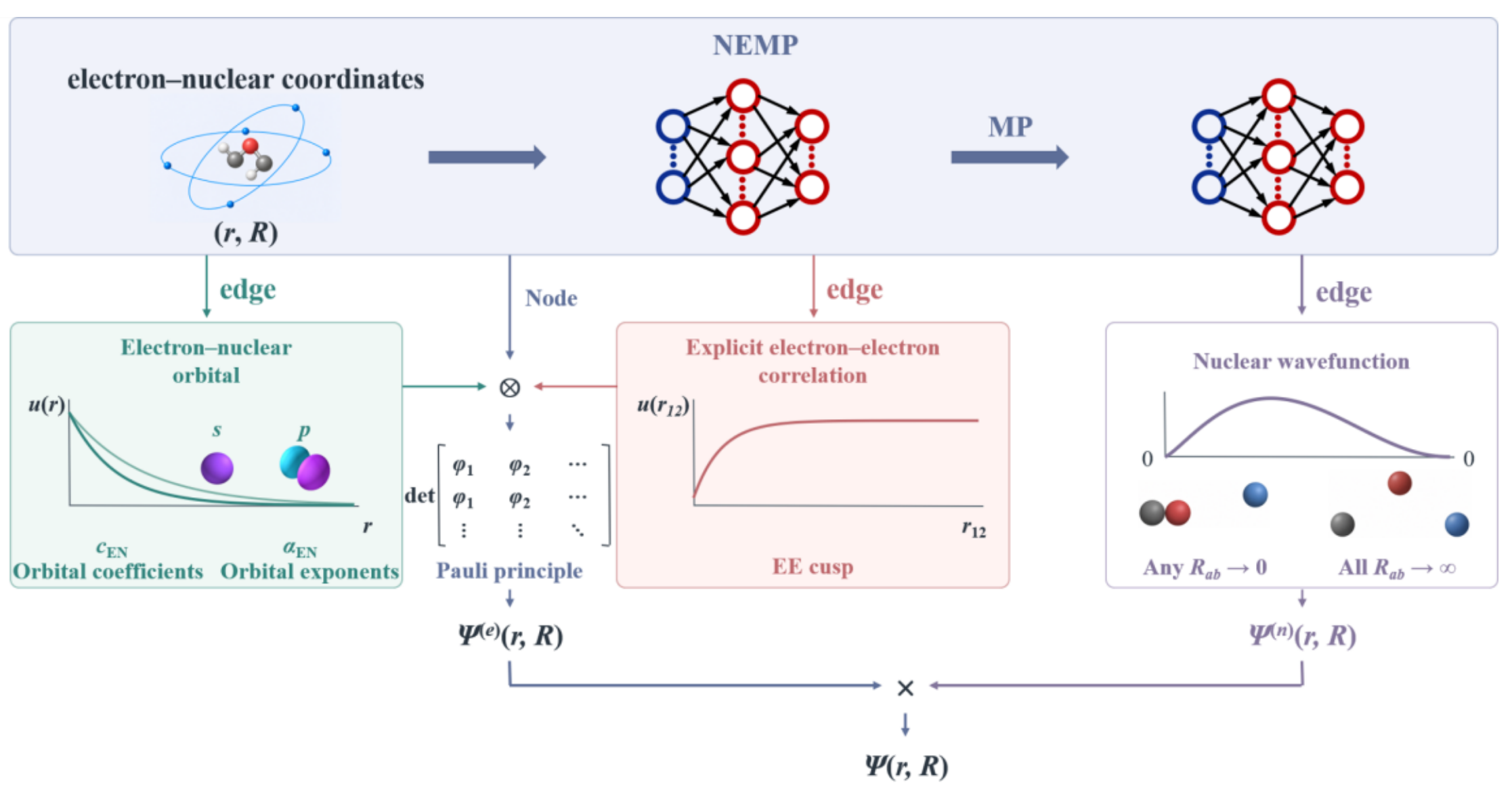

**Fig. 1** Schematic of the configuration-correlated orbital-based wavefunction ansatz in SchrödingerNet.

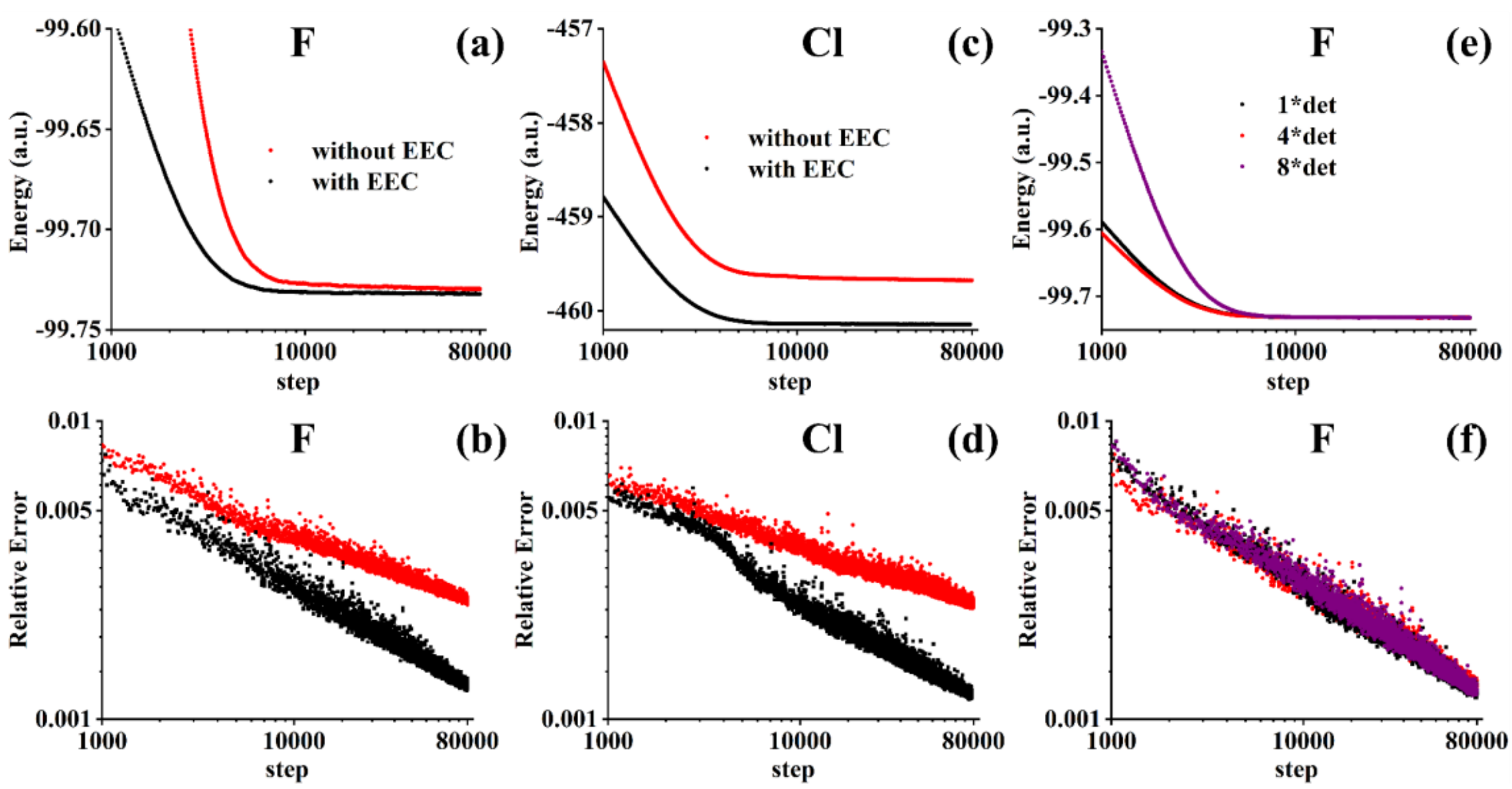

**Fig. 2** Comparisons of the effect of the explicit EEC term and the number of determinants on the SchrödingerNet wavefunction optimization for the F and Cl atoms. Electronic energies and corresponding relative training errors are shown as functions of the optimization step for F in panels **(a)** and **(b)**, for Cl in panels **(c)** and **(d)**, with (black) and without (red) the explicit EEC term, and for F using a single (black), four (red), and eight (purple) Slater determinants in panels **(e)** and **(f)**, respectively. The relative training error is defined as the root-mean-square deviation of local energies from $E_{\mathrm{VMC}}$, divided by $E_{\mathrm{VMC}}$.

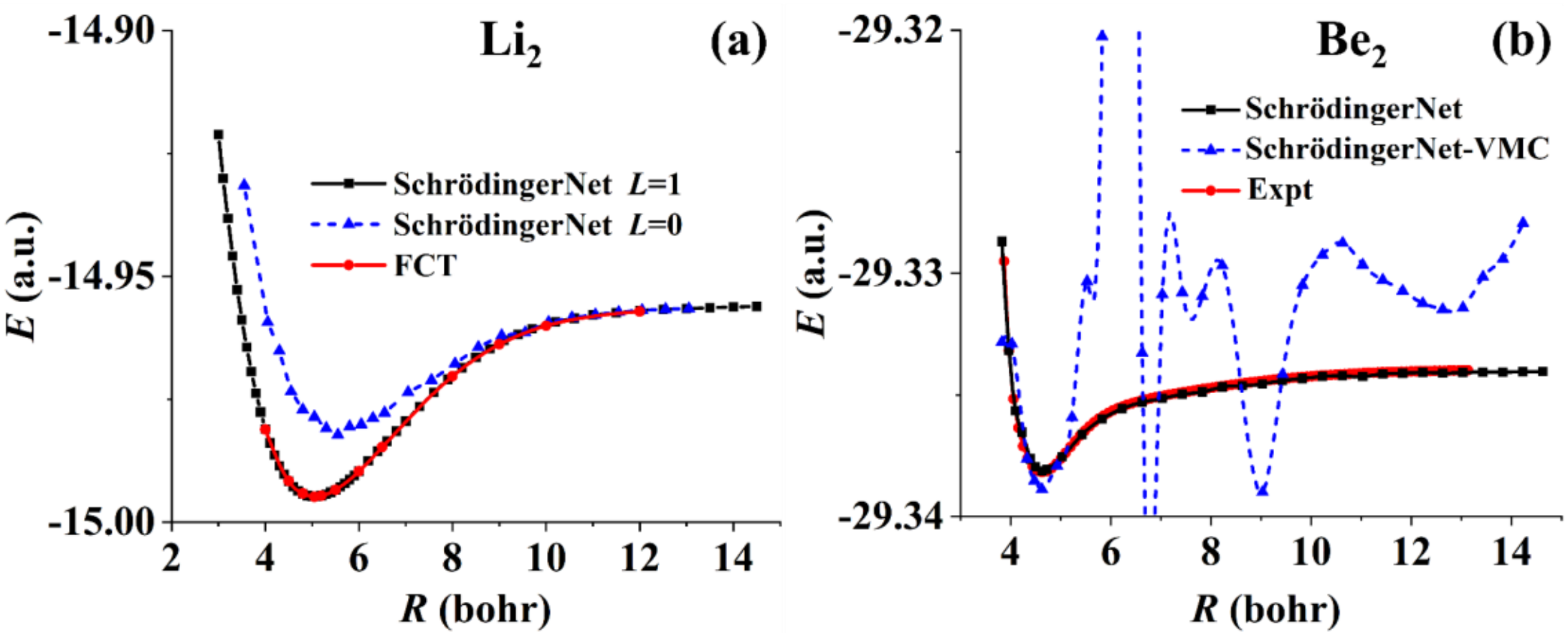

**Fig. 3** **(a)** $Li_2$ electronic energies obtained using SchrödingerNet ($L$=0 and $L$=1), compared with results from those obtained with free complement theory (FCT)[55]. **(b)** $Be_2$ electronic energies obtained using SchrödingerNet with VMC and the hybrid loss function, compared with experimental measurements[60].

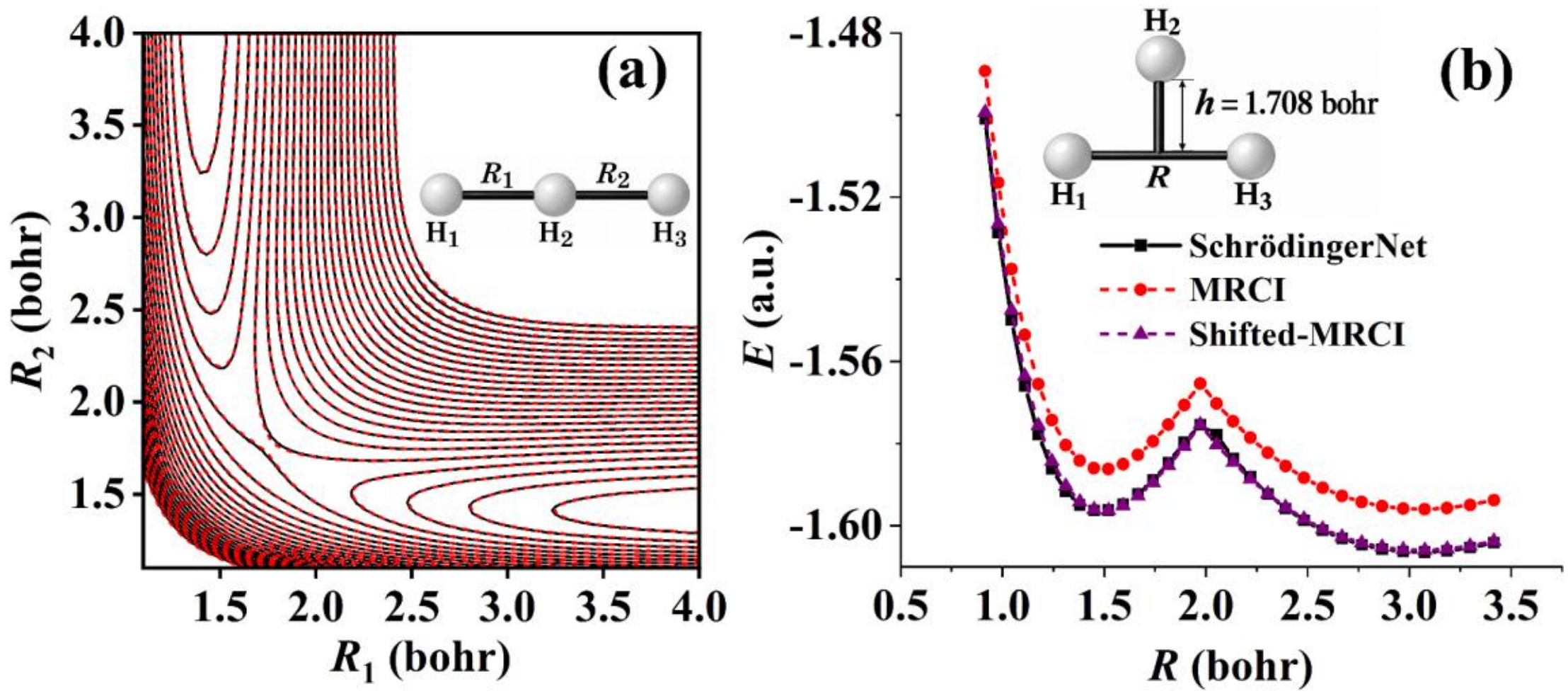

**Fig. 4 (a)** Comparison of the PES obtained using SchrödingerNet (black solid) and the MRCI-based PESs[64] (red dashed) for the H+$H_2$ reaction in the collinear geometry. **(b)** Energy profiles obtained from SchrödingerNet (black), MRCI (red), and Shifted-MRCI (blue) along the H1-H3 distance ($R$) with the vertical distance between H2 to the center of H1-H3 ($h$) fixed at 1.708 bohr, as illustrated in the insert. This path passes through a $D_{3h}$ conical intersection point near $R$=2.0 bohr.

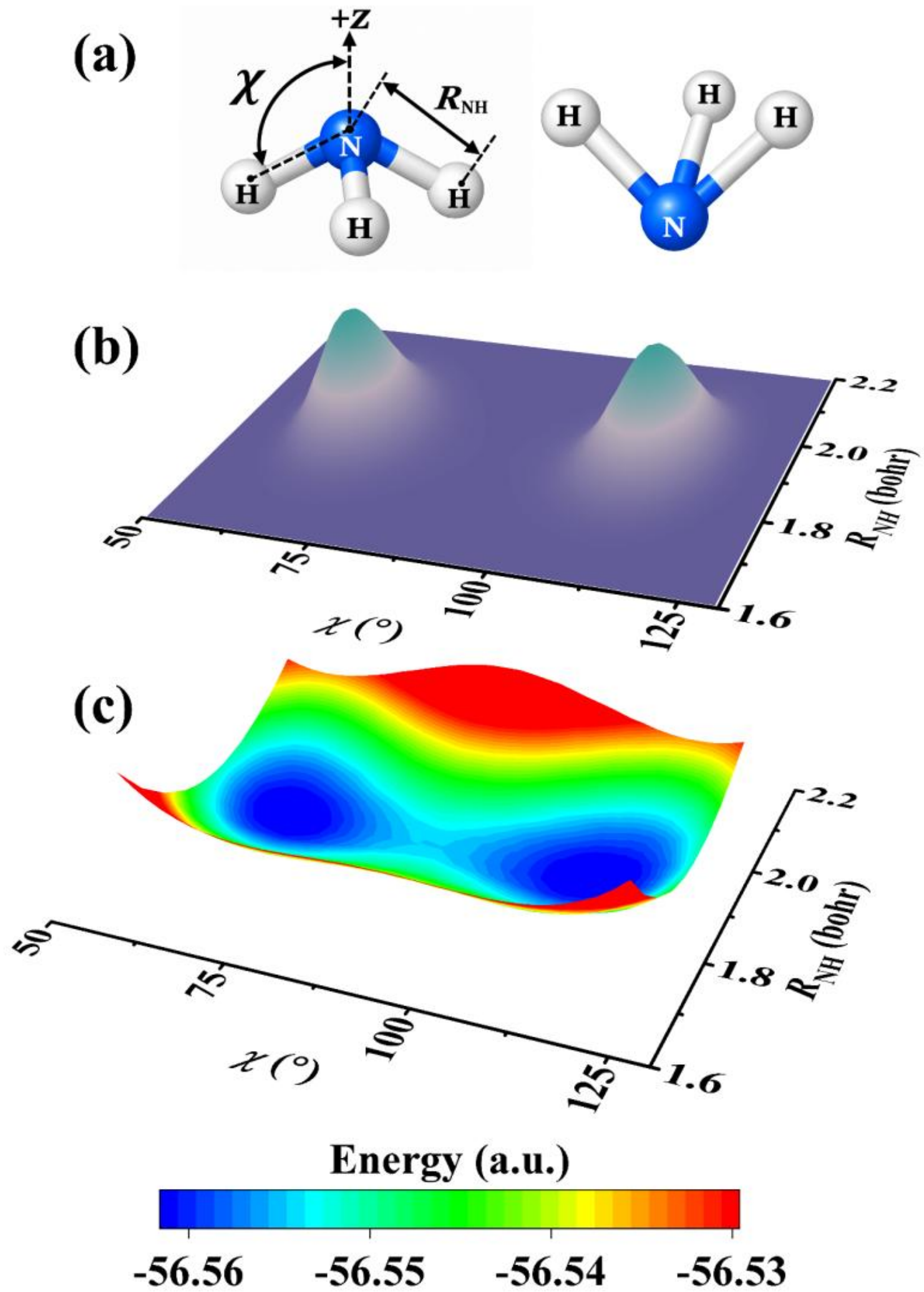

**Fig. 5** **(a)** Definition of the N–H bond lengths ($R_{NH}$) and polar angle ($\chi$) in $NH_3$, measured between an N–H bond and the (+$z$) axis. **(b)** Nuclear probability density obtained from the electron-nuclear wavefunction in the $R_{NH}$–$\chi$ space. **(c)** The corresponding PES. Throughout the scan, N remains fixed at the origin, and the three H atoms retain equal N–H bond lengths and fixed azimuthal angles separated by 120°. The nuclear probability density is estimated by counting $R_{NH}$ and $\chi$ from configurations sampled by VMC with other nuclear degrees of freedom fixed.

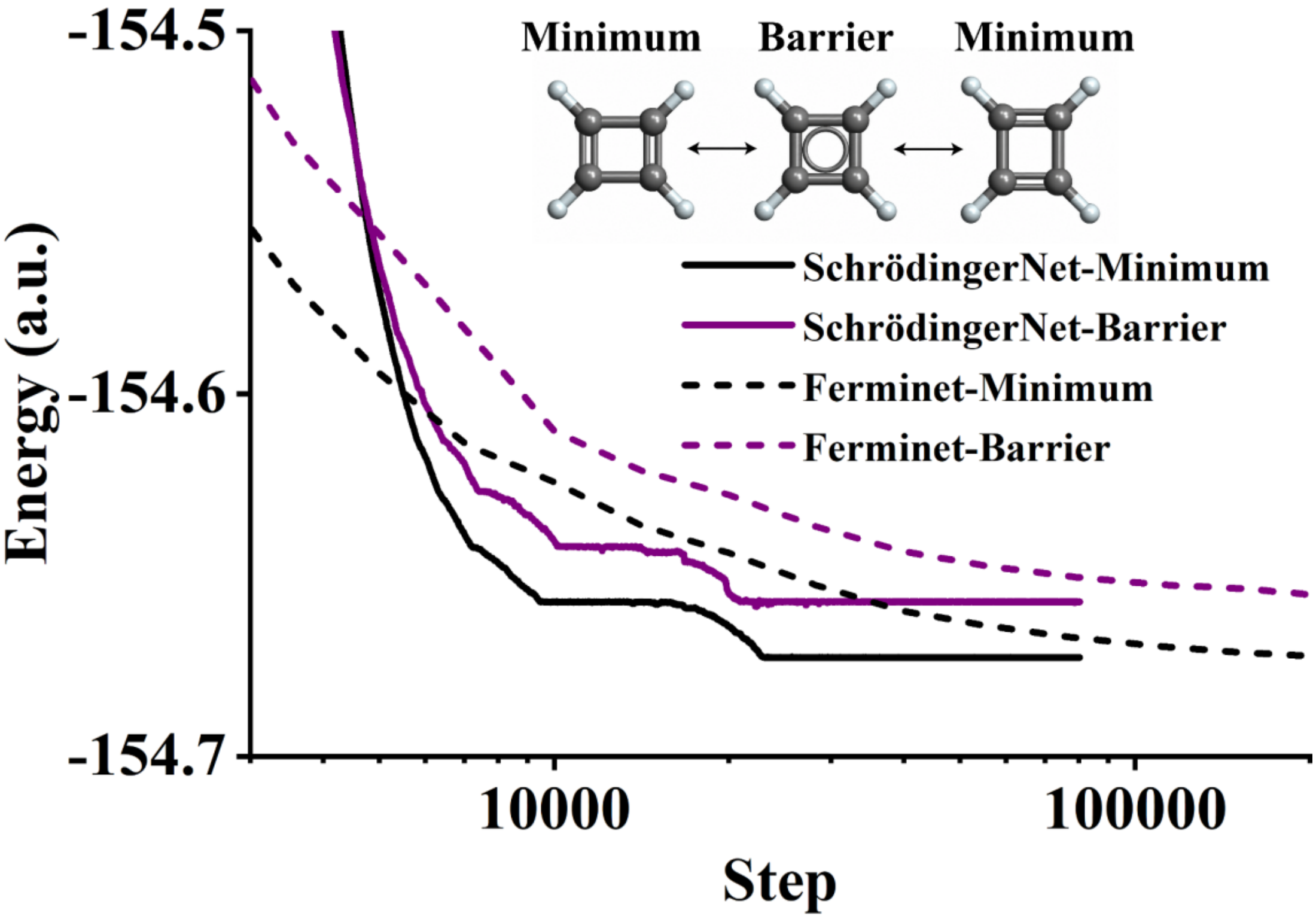

**Fig. 6** Comparison of total-energy learning curves for SchrödingerNet and FermiNet at the minimum-energy and barrier geometries of cyclobutadiene automerization.

**Table I**. Electronic energies of the F and Cl atoms obtained using SchrödingerNet with (with EEC) and without the explicit EEC term (without EEC), compared with the FermiNet results[19, 53]. All energies are reported in hartree.

| Method | SchrödingerNet (with EEC) | SchrödingerNet (without EEC) | FermiNet |
|---|---|---|---|
| F | -99.7330 | -99.7301 | -99.7329 |
| Cl | -460.1445 | -459.6766 | -460.1477 |

**Table II.** Comparison of automerization barrier heights of $C_4H_4$, defined as the energy difference between the square transition state and the rectangular equilibrium structure, as predicted by different methods. All values are reported in kcal/mol.

| Method | CASSCF | CCSDTQ | PauliNet | FermiNet | SchrödingerNet |
|---|---|---|---|---|---|
| Barrier | 11.59 | 9.11 | 9.90 | 10.30 | 9.60 |